\documentclass[aps,pra,superscriptaddress,twocolumn,showpacs,floatfix]{revtex4-2}

\usepackage{amssymb,amsmath,amsfonts,mathrsfs,bm, bbm, graphicx, epsfig, epstopdf}
\usepackage[version=3]{mhchem}
\usepackage{braket}
\usepackage{hyperref}
\usepackage{color}
\usepackage{xcolor}
\usepackage{adjustbox}
\usepackage{csquotes}
\usepackage{float}
\usepackage{colortbl}
\usepackage{siunitx}
\DeclareSIUnit{\nothing}{\relax}
\usepackage{makecell}
\usepackage[T1]{fontenc}
\usepackage{soul}

\date{\today}
\begin{document}
\title{Efficient Pumping of Spectral Holes in a Tm$^\text{3+}$: YAG Crystal for Broadband Quantum Optical Storage}

\author{Yisheng Lei}
\affiliation{Department of Electrical and Computer Engineering and Applied Physics Program, Northwestern University, Evanston, IL 60208, USA}

\author{Zongfeng Li}
\affiliation{Department of Electrical and Computer Engineering and Applied Physics Program, Northwestern University, Evanston, IL 60208, USA}

\author{Mahdi Hosseini}
\thanks{Corresponding author}
\email{mh@northwestern.edu}
\affiliation{Department of Electrical and Computer Engineering and Applied Physics Program, Northwestern University, Evanston, IL 60208, USA}
\affiliation{Elmore Family School of Electrical and Computer Engineering, Purdue University, West Lafayette, Indiana 47907, USA}

\begin{abstract}
Quantum memory devices with high storage efficiency and bandwidth are essential elements for future quantum networks. Here, we report a storage efficiency greater than 28\% in a Tm$^\text{3+}$: YAG crystal in elevated temperatures and without compromising the memory bandwidth. Using various pumping and optimization techniques, we demonstrate multi-frequency window storage with a high memory bandwidth of 630 MHz. Moreover, we propose a general method for large-bandwidth atomic-frequency memory with non-Kramers rare-earth-ion (REI) in solids enabling significantly higher storage efficiency and bandwidth. Our study advances the practical applications of quantum memory devices based on REI-doped crystals.  
\end{abstract}

\maketitle{}

\section{Introduction}
Networked quantum systems consisting of quantum computers, atomic clocks, and other quantum sensors can greatly exceed their individual capabilities \cite{kimble2008quantum, wehner2018quantum}. To establish long-distance quantum networks, quantum repeaters are indispensable elements \cite{briegel1998quantum}. Quantum repeaters based on quantum memories were first proposed \cite{duan2001long, simon2007quantum}, which require less technological complexities compared to other types of quantum repeater architectures \cite{muralidharan2016optimal, azuma2023quantum}. Toward this technical path, absorptive quantum memory devices with high storage efficiency, large memory bandwidth, long storage time, and multimode capacity are under active development \cite{lei2023quantum}. Rare-earth ions in solids are attractive platforms for realizing these functions due to their long coherence times \cite{zhong2015optically} and their properties for optical integration \cite{zhou2023photonic}. Tm$^\text{3+}$ ions have optical transition wavelengths close to that of rubidium atoms; a platform under active investigation for neutral-atom-based quantum computing \cite{bluvstein2024logical}. Thus, Tm$^\text{3+}$ ions in solids may be used for distributed quantum computing or building a hybrid quantum repeater \cite{gu2024hybrid}. To date, quantum storage with Tm$^\text{3+}$ ions in various host crystals, including YAG \cite{chaneliere2010efficient, davidson2020improved}, LiNO$_\text{3}$ \cite{sinclair2014spectral, askarani2020entanglement} and YGG \cite{thiel2014tm, askarani2021long}, has been demonstrated. Tm$^{3+}$ ions can be integrated with Lithium-Niobate-on-Insulator for storage \cite{dutta2023atomic}, spectral filter \cite{zhao2024cavity} or studies of collective behavior \cite{pak2022long}. In particular, Tm$^{3+}$ ions in YAG crystals have a relatively long optical coherence time about 100~$\mu$s, a ground-state lifetime over 1~s at around 1~K, and a high branching ratio of 0.25 \cite{louchet2007branching}, making them a good candidate for spectral tailoring and photon storage.  Although memory efficiency can improve when operating at dilution refrigerator temperatures or using monolithic cavity crystals \cite{davidson2020improved}, operation at elevated temperatures and in free space is desirable to achieve scalability and high bandwidth, respectively. We have recently shown that algorithmic optimization and a multi-path configuration can enhance the storage efficiency of Tm$^\text{3+}$: YAG memories by more than a factor of six \cite{lei2024algorithmic}. Here, we report a more advanced quantum memory in a Tm$^\text{3+}$: YAG crystal with record efficiency and bandwidth at 3.5~K. We effectively implement various pumping sequences to demonstrate efficient, broadband, and multi-frequency storage. We also propose a novel pumping scheme capable of extending the memory bandwidth by more than an order of magnitude.

\begin{figure*}[!t]
\centerline{\includegraphics[width=1.9\columnwidth]{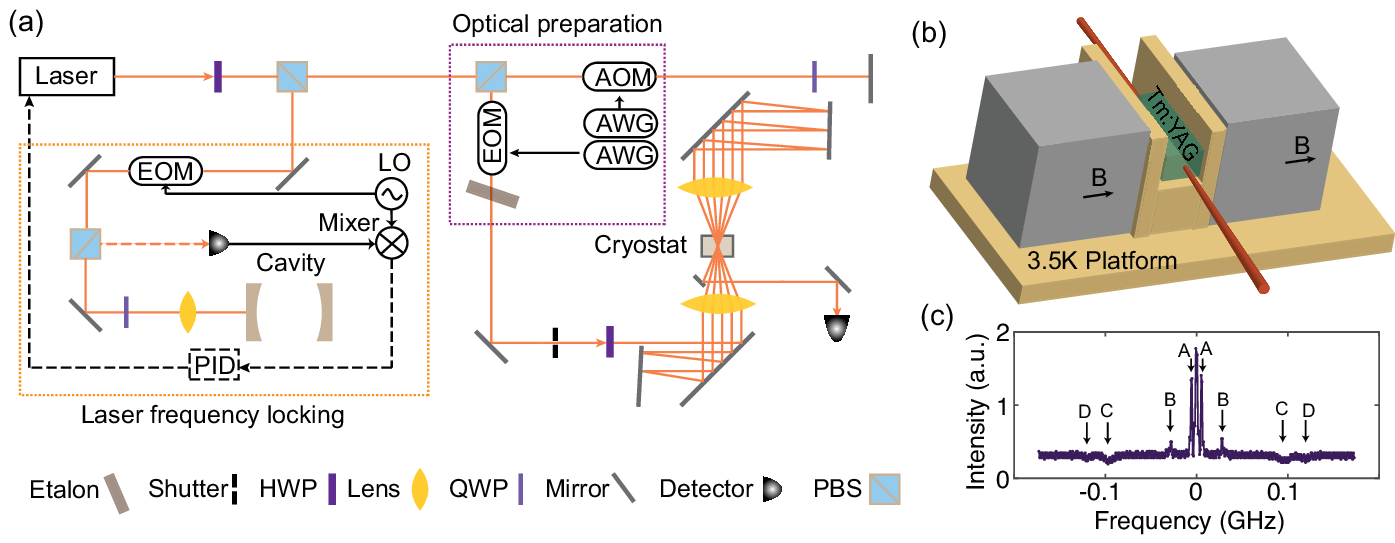}}
	\caption{(a) Experimental setup. Laser frequency locking was implemented using Pound–Drever–Hall (PDH) technique \cite{black2001introduction} and a PID controller shown in the yellow-dotted box. Laser beam (pump and signal) routed through the crystal six times. A double-pass acousto-optic modulator (AOM), electro-optic modulator (EOM), a two-channel arbitrary waveform generator (AWG) and etalon were used to implement different pumping schemes. (b) Schematics of the memory setup with copper platform attached to the cold head of the cryostat with a base temperature of 3.5~K; Tm$^\text{3+}$: YAG crystal is placed on top of the platform and two permanent block magnets are fixed in parallel. (c) Spectrum of the Tm$^\text{3+}$ ion ensemble by spectral hole burning showing a spectral hole (center peak), side holes (A, B) and anti-holes (C, D).}
	\label{Fig1}
\end{figure*}

\section{Experiment}
The optical setup is shown in Fig. \ref{Fig1}(a). The laser (Toptica DL Pro) used for the  experiment has a nominal linewidth of 1~MHz. A small portion of the laser power is used for laser frequency stabilization. The stabilizing cavity consists of two concave mirrors with high reflectivity of 99.9\% (finesse $\sim$ 3000) placed inside an ultra-low-expansion glass, which is temperature controlled to tune the laser frequency. The laser linewidth after frequency locking to the stabilizing cavity is estimated to be below 100~kHz. The main power of the laser is sent to an acousto-optic modulator (AOM) with a double-pass configuration, with a modulation bandwidth of 50~MHz. The laser beam is typically routed such that to pass through the crystal six times, which is achieved by two lenses and a few mirrors. To increase the frequency tuning range for broadband storage, an electro-optic modulator (EOM) and etalon (for multi-frequency storage) are placed after the AOM. 

A Montana cryostat provides a base temperature of 3.5~K. A custom-designed platform, which can properly fix the crystal and block magnets while maintaining the cooling power of the crystal, is attached to the cold head (see Fig. \ref{Fig1}(b)). The Tm$^{3+}$: YAG crystal with dimensions $4\times5\times10$ mm has a doping concentration of 0.1\%. The laser beam propagates parallel to the $\langle$110$\rangle$ axis of length 10~mm with linear polarization. The optical transition $^{\text{3}}$H$_{\text{4}}$ $\leftrightarrow$ $^{\text{3}}$H$_{\text{6}}$ occurs at a wavelength of $\lambda=$793.373~nm. The two permanent block magnets produce a magnetic field of 4500~G along the $\langle$001$\rangle$ axis, lifting the degeneracy of Tm$^{3+}$ ions with a nuclear spin of $1/2$. One of the two ground states can serve as a shelving state for spectral preparation. The two-pulse echo measurements estimate the optical coherence time of the Tm$^{3+}$ ion ensemble to be 38~$\mu$s. The measurement of the spectral hole depth after pumping shows a double-exponential decay with a fast decay time of 10~ms and a slow decay time of 170~ms, corresponding to the intermediate level $^{\text{3}}$F$_{\text{4}}$ lifetime and the ground state lifetime \cite{sinclair2021optical}, respectively. The spectral hole linewidth of the Tm$^\text{3+}$ ion ensemble is measured to be 0.5~MHz, mainly limited by spectral diffusion, which can be reduced by using lower temperatures. The hole linewidth measurement is performed at low pumping powers, so the power broadening is negligible. There are two classes of ions observed: one class has an excited state splitting of 6~MHz, for which no anti-hole was observed; another class has an excited state splitting of 27~MHz and a ground state splitting of 125~MHz, as seen in Fig. \ref{Fig1}(c). This is consistent with previous spectral measurements of Tm$^{3+}$: YAG crystal \cite{de2006experimental}.

\section{Results}
\subsection{Efficient Storage}

\begin{figure*}[!t]
\centerline{\includegraphics[width=1.85\columnwidth]{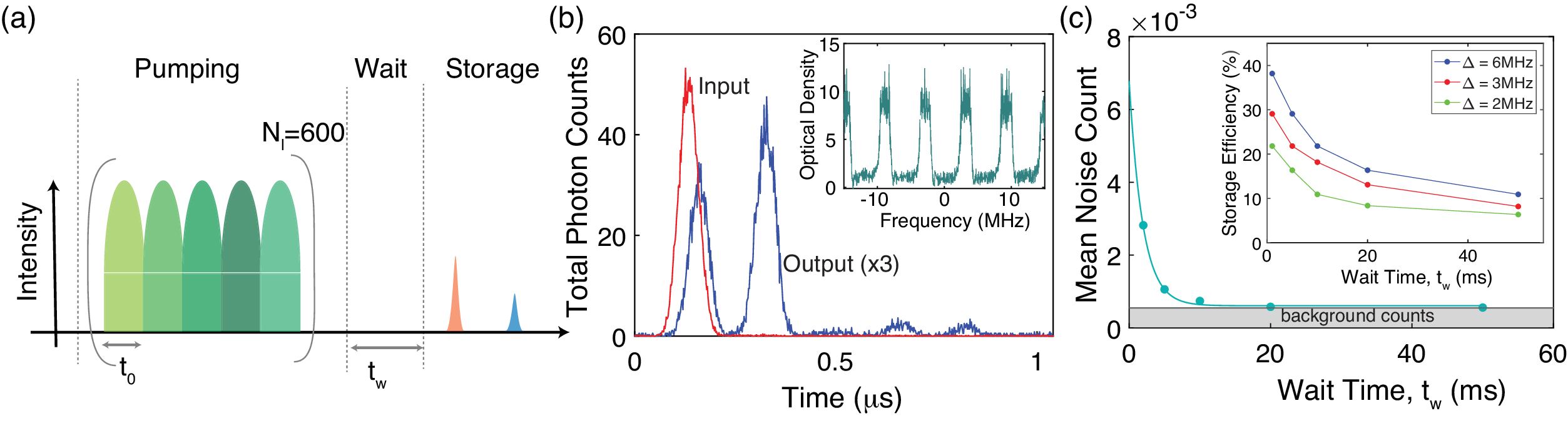}}
	\caption{(a) Pumping sequence by sweeping the input RF frequency to the AOM. (b) Total single photon counts of 10000 storage events performed at 5~ms after spectral preparation. The red curve is input pulse for storage, and the blue curve represents transmission and echo pulses. The inset shows the optical density of the Tm$^\text{3+}$ ion ensemble measured at 5~ms after spectral preparation for AFC $\Delta$ = 6~MHz. (c) Average noise counts during a time window of 100~ns for different wait times after spectral pumping. There is a constant background counts from the pump laser leakage (5.5$\times$10$^{-4}$ counts) and detector dark counts (1$\times$10$^{-5}$ counts) during a time window of 100~ns. The inset figure shows storage efficiency for different AFC $\Delta$ with different wait times after spectral pumping.}
	\label{Fig2}
\end{figure*}

The atomic frequency comb (AFC) is a photon-echo-based protocol for optical storage, which is based on spectral tailoring and atomic rephasing \cite{afzelius2009multimode}. An atomic medium with high-finesse comb profiles properly tailored into its absorption spectrum can enable efficient storage at high optical densities. In our experiment, the laser beam is routed to pass through the crystal six times increasing the effective optical density to $\sim$12. For an effective spectral preparation, we use adiabatic pumping pulses shaped with a secant hyperbolic function and frequency chirped with a tangent hyperbolic function, which can create frequency combs with square profiles \cite{bonarota2010efficiency, jobez2016towards}. Each comb is pumped with a train of these pulses with duration $t_0$, and the pulse sequence is repeated $N_l$ times. After a wait time of $t_w$, designed to avoid spontaneous emission while excited atoms decay, storage pulses with duration of $\sim$80~ns are sent to the crystal. The bandwidth ($\Delta_p$),  intensity profile of the pumping pulses, $t_0$ and $N_l$ (see Fig. \ref{Fig2}(a)) are optimized to obtain the highest storage efficiency. For $\Delta$ = 6~MHz, the optimized parameters are $N_l$ = 600, $t_0$ = 0.15~ms and $\Delta_p$ = 4.2~MHz. Quantum storage with single-photon-level storage pulses are performed for $t_w=$5~ms after the spectral pumping (see Fig. \ref{Fig2}(b), and the storage efficiency is determined to be 28.5$\pm$0.2\%. The corresponding optical density of the medium for $t_w=$5~ms is shown in the inset figure of Fig. \ref{Fig2}(b). Storage efficiency and average noise counts for different waiting times are also recorded, as shown in Fig. \ref{Fig2}(c).  If the input pulses have one photon on average, the signal-to-noise ratio at $t_w$ = 5~ms is estimated to be $\sim$270 for a photon collection window of 100~ns. The noise photons are due to the spontaneous emission of the atoms remaining in the optically excited state after spectral pumping, constant leakage from the laser, and also dark counts of the single-photon detector. The same experiments are repeated for $\Delta$ = 3~MHz \& 2~MHz, corresponding to longer storage times.

For atomic frequency combs with square comb profiles created by the adiabatic pulses, the storage efficiency can be estimated using the following equation \cite{bonarota2010efficiency, jobez2016towards},
\begin{equation}\label{Eqa1}
\eta = \frac{d^2}{F^2}\text{exp}(-\frac{d}{F})\text{sinc}^2(\frac{\pi}{F})\text{exp}(-d_0), 
\end{equation}
where $d$ is the optical depth (OD) of the atomic medium before optical pumping, $F$ is the finesse of the combs, and $d_0$ is the OD corresponding to the background absorption after spectral tailoring. For $\Delta$ = 6~MHz, the finesse is estimated to be $\sim$4.5. $d_0$ is estimated to be $\sim$0.4. The theoretical storage efficiency is calculated to be 30.4\%. This agrees with our experimental results, considering that any imperfection in the comb shape leads to stronger self-dephasing, lowering storage efficiency. For square AFC profiles, it can be shown that an optimal finesse maximizes the first echo and suppresses the second and fifth echoes \cite{Li2025}. The storage echoes in Fig. \ref{Fig2}(b) confirm that the AFC is operating under optimal finesse conditions.

\subsection{Multi-frequency-bin Storage}
\begin{figure*}[!t]
\centerline{\includegraphics[width=1.5\columnwidth]{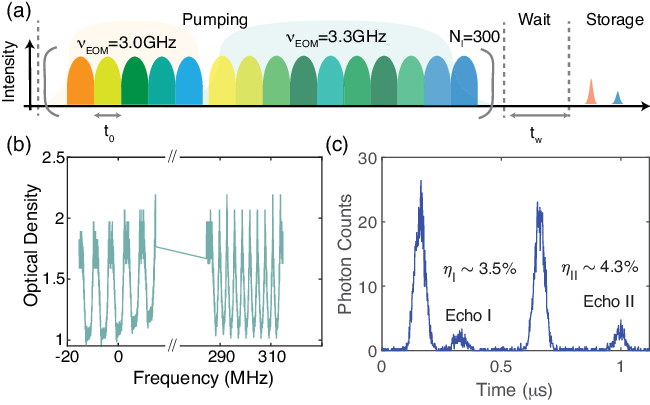}}
	\caption{(a) Pumping sequence by sweeping the input RF frequency to the AOM and inputting two different RF frequencies to the EOM. (b) Optical density of two frequency windows separated by 300~MHz measured at 5~ms after spectral preparation for AFC $\Delta_1$ = 6~MHz and $\Delta_2$ = 3~MHz respectively. (c) Single photon measurements of two input storage pulses with frequencies separated by 300~MHz and 10000 storage events are performed.}
	\label{Fig3}
\end{figure*}

The spectral tailoring with AOM has a limited bandwidth of $\sim$ 100~MHz. To avoid this limitation, an EOM and an etalon (see Fig. \ref{Fig1}(a)) are used to prepare the AFC with two frequency windows separated by 300~MHz. The etalon is monitored by a temperature controller that adjusts the resonant frequency of the etalon to match the absorption spectrum of Tm$^\text{3+}$ ions. The etalon has a bandwidth of 500~MHz. The extinction ratio of the EOM was found to be $\sim$20, which was measured by driving the EOM with a two-tone RF signal ($\nu_{\text{EOM}}$) with frequencies $f_1$ = 3.0~GHz \& $f_2$ = 3.3~GHz (see Fig. \ref{Fig3}(a)). The choice of 300~MHz frequency separation ensures higher optical power for spectral preparation. Due to the limited laser power, a Tm$^\text{3+}$ ion ensemble with a lower optical density should be used for storage. Therefore, the laser beam is routed to pass through the crystal only once. An AFC with $\Delta_1$ = 6~MHz is prepared around the center frequency $f_1$ = 3.0~GHz, and an AFC with $\Delta_2$ = 3~MHz is prepared at $f_2$ = 3.3~GHz, which correspond to storage times of 167~ns and 333~ns respectively. 

The pumping parameters are optimized to obtain the highest storage efficiency, returning $N_l$ = 300, $t_0$ = 0.1~ms and $\Delta_{p1}$ = 4.2~MHz \& $\Delta_{p2}$ = 1.8~MHz. The AFC spectrum measured at 5~ms after spectral preparation is shown in Fig. \ref{Fig3}(b). Single-photon-level storage is performed with wait time $t_w$ = 5~ms, and the storage efficiency is estimated to be $\sim$3.5\% for $\Delta_1$ = 6~MHz and $\sim$4.3\% for $\Delta_2$ = 3~MHz respectively, shown in Fig. \ref{Fig3}(c). As mentioned earlier, the efficiency in this case is limited by the laser power.

\subsection{Broadband Storage}

\begin{figure*}[!t]
\centerline{\includegraphics[width=1.85\columnwidth]{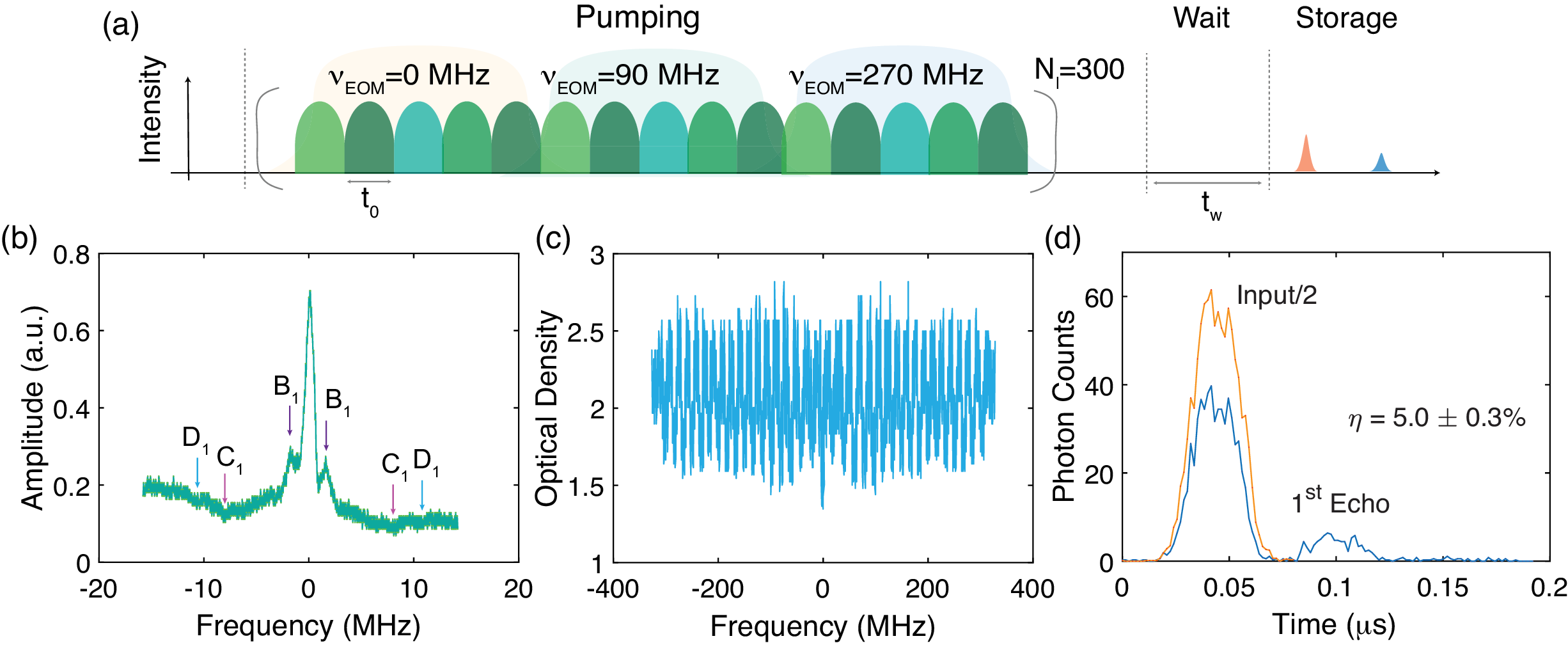}}
	\caption{(a) Pumping sequence by sweeping the input RF frequencies to the AOM and inputting three different frequencies to the EOM. (b) Spectrum of the Tm$^\text{3+}$ ion ensemble with a magnetic field of 370~G showing spectral hole, side holes (B$_\text{1}$) and anti-holes (C$_\text{1}$, D$_\text{1}$). (c) Optical density of the Tm$^\text{3+}$ ion ensemble with bandwidth of 630~MHz and AFC $\Delta$ = 18~MHz. (d) Single photon measurements of storage pulses with durations of 25~ns.}
	\label{Fig4}
\end{figure*}

The intrinsic pumping scheme creates AFC by transferring atoms from one frequency window to an adjacent window without reducing the overall optical density \cite{askarani2021long}. In this case, the finesse of the AFC is intrinsically $\sim$2. Referring to Eq. \ref{Eqa1}, storage medium with an optical density of 2 after spectral tailoring gives the best storage efficiency. Therefore, for implementing the intrinsic pumping scheme, we route the laser beam once through the crystal, which corresponds to a maximum optical depth of 2.2, where also
\begin{equation}
    \Delta_g - \Delta_e \simeq \frac{\Delta_{AFC}}{2}.
\end{equation}
In this way, the entire inhomogeneously broadened absorption window can be used for optical storage, which can realize broadband quantum storage \cite{askarani2021long}. The background absorption after spectral preparation can be reduced by matching the spectral holes and anti-holes with AFC troughs and valleys, which can further enhance memory efficiency. For realizing this pumping scheme, the magnets are moved further apart to reduce the magnetic field strength at the crystal to be about 370~G. Multiple EOM tones were used together with the AOM-frequency tuning to create a broadband AFC (Fig. \ref{Fig4}(a)). The spectrum after a single spectral hole burning is shown in Fig. \ref{Fig4}(b). The center of anti-holes and the center of spectral holes are separated by 9~MHz, which means AFC $\Delta$ should be set to be 18~MHz, corresponding to a storage time of 55.6~ns. To create the AFC, the AWG is used to drive the EOM with three distinct frequencies and amplitudes to create seven tones separated by 90~MHz (three equal-amplitude sidebands on each side of the carrier). Using the AOM, we set $\Delta_p$ = 8.2~MHz for each comb with $t_0$ = 0.1~ms. This gives a comb linewidth of $\sim$9~MHz considering the additional spectral broadening. In total, a memory bandwidth of 630~MHz (35 comb lines separated by 18~MHz) is created and its optical density spectrum is shown in Fig. \ref{Fig4}(c). Different combinations of RF frequencies and amplitudes for the EOM and AOM are studied to find an optimal sequence for the AFC preparation. After pumping is completed, we add a 5~ms of wait time to allow atoms to decay back to the ground states, and then carry out the storage experiment by sending 25~ns weak coherent pulses. Single-photon-level storage is performed and a memory efficiency of 5.0$\pm$0.3\% was observed, as shown in Fig. \ref{Fig4}(d). To create such broadband AFC memories, more atoms have to be pumped to the other ground state, which requires higher optical power (above 10~mW) leading to crystal heating. As a result, longer spectral preparation time is required, which works well only if the ground-state lifetime is long enough. In this experiment, the memory efficiency is mainly limited by the short ground-state lifetime, which leads to inefficient spectral preparation.


\subsection{Commensurate Intrinsic Pumping}
We now discuss a new pumping scheme that enables more efficient AFC pumping to create broadband memories. 
\begin{figure}[!h]
\centerline{\includegraphics[width=0.95\columnwidth]{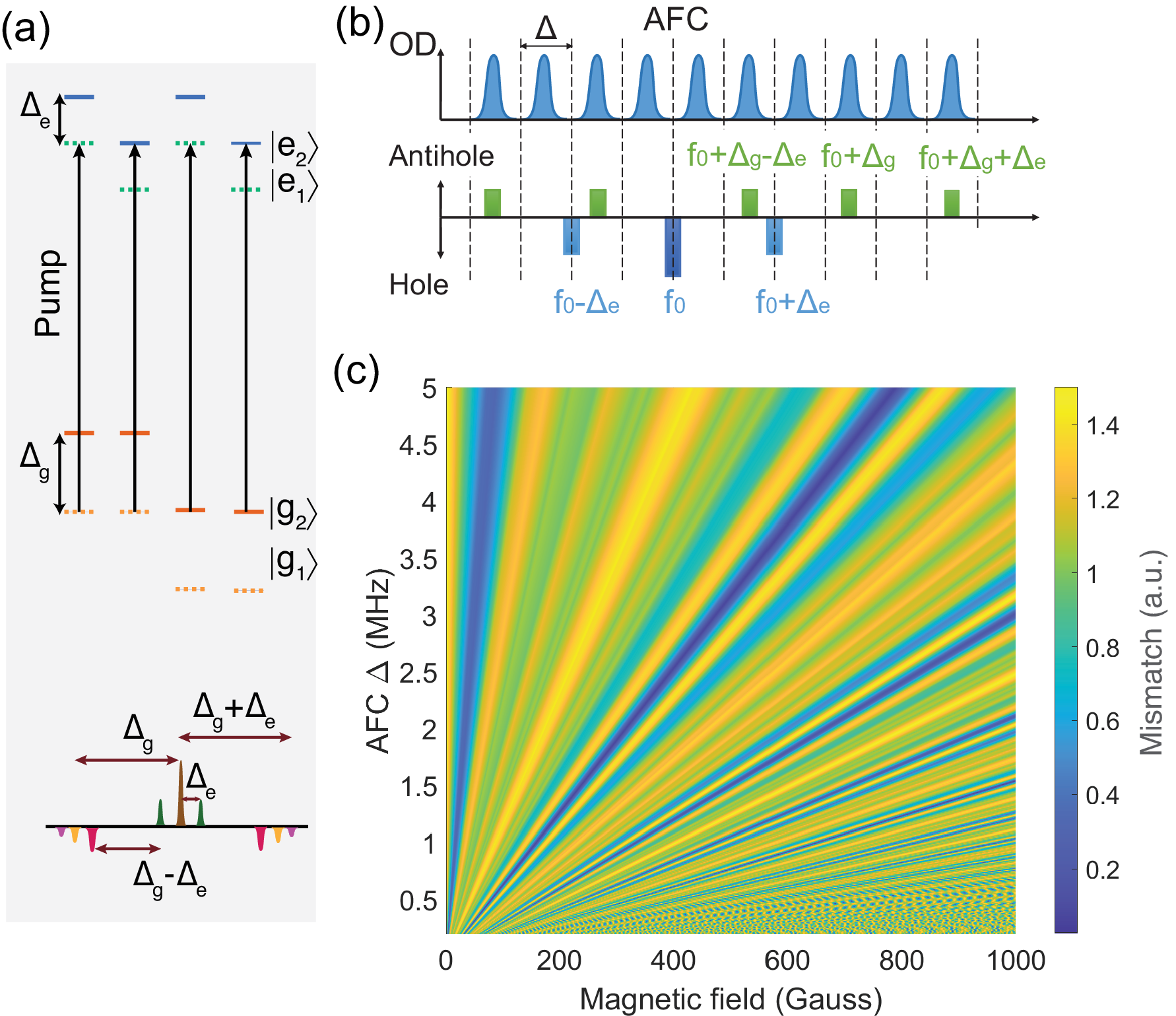}}
	\caption{(a) Spectral hole burning for Tm$^\text{3+}$ ions in YAG crystal with inhomogeneous broadening: one spectral hole, two side holes and six anti-holes are shown. (b) The anti-holes and holes match the AFC spacing in the intrinsic pumping scheme. The excited and ground states of the ions are doublets, with a splitting of $\Delta_e$ and $\Delta_g$, respectively. (c) The degree of mismatch calculated as function of applied magnetic field and AFC spacing, $\Delta$. The mismatch is found by the sum of the distances between $n_i$ (see the text) and an integer. The splitting are $\mu_e = 0.006$ MHz/G and $\mu_e = 0.0285$ MHz/G \cite{de2006experimental}.}
	\label{Fig5}
\end{figure}

Let's consider the Tm$^\text{3+}$ ions in YAG crystal, where the ground state splitting is 4.75 times of the excited state spitting. When a magnetic field on order of 100~G is applied, the side holes overlap with the center spectral hole and the anti-holes overlap with each other at each side, due to the spectral broadening. In this scenario, the ``commensurate intrinsic pumping" can be achieved. This is because the spin-conserved transitions have higher dipole strengths compared with the spin-crossed transitions, which make the anti-holes at $\pm\Delta_g$ smaller and the anti-holes at $\pm$($\Delta_g$+$\Delta_e$) much smaller, as shown in Fig. \ref{Fig5}(a). 

Comparing with the other spectral tailoring schemes, the basic idea of the proposed commensurate intrinsic pumping is to effectively leverage all the anti-holes and holes employed as the AFC peaks and valleys, respectively. Thus, such an approach enables ultra-wide bandwidth (up to the whole inhomogeneous broadening of atoms), and utilizes all ions in this spectral range. For REIs with two excited and two ground states in the inhomogeneously broadened spectrum, a single frequency ($f_0$) laser will burn three holes at $f_0 + s\cdot\Delta_e$, and six anti-holes at $f_0 \pm \Delta_e + s\cdot\Delta_g$, where $s=0,\pm1$. The commensurate intrinsic pumping scheme will hold when all of the anti-holes and holes match the AFC spacing, as shown in Fig. \ref{Fig5}(b). In this way, a series of holes can be pumped with width $\Delta/2$ at every $\Delta$ interval to form an AFC with ultimate bandwidth.
In this regime, the splittings $\Delta_{e,g}$ and the AFC spacing $\Delta$ satisfy:

\begin{equation}
    \begin{split}
        \Delta_e &= n_1\times\Delta,\\
        \Delta_g &= (n_2 + \frac{1}{2})\times\Delta,\\
        |\Delta_g - \Delta_e | &= (n_3 + \frac{1}{2})\times\Delta,\\
        |\Delta_g + \Delta_e | &= (n_4 + \frac{1}{2})\times\Delta.
    \end{split}
\end{equation}
where $n_i$ are integers. Considering the splittings proportional to the applied magnetic field, i.e. $\Delta_{e,g} = \mu_{e,g}\cdot B$, there are two degrees of freedom to reach the above intrinsic conditions: the field $B$ and the storage time $\frac{1}{\Delta}$.
We have performed simulations for Tm$^\text{3+}$: YAG crystal. To quantify the degree of mismatch, we calculate four values ( $n_i$ ) that satisfy the four equations above, with a given magnetic field B and $\Delta$. Each value's distance from the nearest integer is at most 0.5. The mismatch is defined as the sum of these distances, divided by 2 (or equivalently, by $0.5\times 4$), where 0 correspond to the maximum matching. This quantity is plotted in Fig. \ref{Fig5}(c) with respect to the varying magnetic field and storage time.
The result indicates that we can always choose a field to reach a given storage time. For example, when 250~ns storage time is required, a field of 630~G could gives a 96.5\% match; for a 1~$\mu$s storage, a field of 158~G and 647.5~G give a 96\% and 98.8\% match respectively. Similar simulations can be applied to other non-Kramers ions in various types of crystals for intrinsic pumping, which enable memory bandwidth with a few orders of magnitude higher than the previous experiments. To efficiently implement this scheme, the spectral hole linewidth of the REI has to be narrow enough compared with the desired AFC $\Delta$. In our experiment, the spectral hole linewidth is limited by spectral diffusion and can be reduced by operating at lower temperatures to implement the commensurate intrinsic pumping. 


\section{Discussions}
In our experiments, the ground-state lifetime of Tm$^{3+}$ ions limit the performance of the spectral tailoring. Lower temperatures can prolong the ground-state lifetime and reduce the spectral hole linewidth. When the ground state lifetime is above 500~ms, the optical depth of background absorption can be reduced to below 10\%, which will give a storage efficiency of 40\%. For AFC memory with multiple frequency windows, a higher laser power can prepare an AFC with much lower background absorption, further enhancing the storage efficiency. Using the pumping scheme discussed above, multiple AFC frequency windows can be tailored for multiplexed storage, coherent manipulation of quantum information, and storage of high-dimensional frequency qubits \cite{yang2018multiplexed, li2024efficient}. A longer ground-state lifetime, achievable at lower temperatures, enables the preparation of high-efficiency memories with a bandwidth of ~2 GHz and storage efficiency exceeding 30\% by intrinsic pumping. In addition, the commensurate intrinsic pumping scheme proposed above can be used for a variety of non-Kramers ions in different types of crystals. Cavity-impedance matching conditions give another way to further boost memory efficiency but in the expense of limiting the memory bandwidth \cite{afzelius2010impedance, moiseev2010efficient}.

\section{Conclusions}
We have implemented various pumping schemes to create efficient and broadband AFC in a Tm$^{3+}$-doped YAG crystal. A storage efficiency of 28.5$\pm$0.2\% for a memory bandwidth of 30~MHz were reported. This record-high efficiency was achieved without using a monolithic crystal cavity or a dilution refrigerator. A pumping scheme for storage of multiple spectral modes was developed and an effective pumping scheme for large-bandwidth storage (630~MHz) was also demonstrated. Lastly, we proposed the commensurate intrinsic pumping scheme to reach even higher bandwidth storage. \\

\section{Acknowledgments}
 We would like to thank Mehrdad M. Sourki for helping to machine the platform inside the cryostat. We acknowledge the support from National Science Foundation Career Award number 2410198, and U.S. Department of Energy, Office of Science, Office of Advanced Scientific Computing Research, through the Quantum Internet to Accelerate Scientific Discovery Program under Field Work Proposal 3ERKJ381.

\section{Disclosures}
The authors declare no conflict of interest.\\

\section{Data availability}
Data underlying the results presented in this paper are not publicly available at this time, but may be obtained from the authors upon reasonable request.

\bibliography{sample}{}

\begin{thebibliography}{34}%
\makeatletter
\providecommand \@ifxundefined [1]{%
 \@ifx{#1\undefined}
}%
\providecommand \@ifnum [1]{%
 \ifnum #1\expandafter \@firstoftwo
 \else \expandafter \@secondoftwo
 \fi
}%
\providecommand \@ifx [1]{%
 \ifx #1\expandafter \@firstoftwo
 \else \expandafter \@secondoftwo
 \fi
}%
\providecommand \natexlab [1]{#1}%
\providecommand \enquote  [1]{``#1''}%
\providecommand \bibnamefont  [1]{#1}%
\providecommand \bibfnamefont [1]{#1}%
\providecommand \citenamefont [1]{#1}%
\providecommand \href@noop [0]{\@secondoftwo}%
\providecommand \href [0]{\begingroup \@sanitize@url \@href}%
\providecommand \@href[1]{\@@startlink{#1}\@@href}%
\providecommand \@@href[1]{\endgroup#1\@@endlink}%
\providecommand \@sanitize@url [0]{\catcode `\\12\catcode `\$12\catcode `\&12\catcode `\#12\catcode `\^12\catcode `\_12\catcode `\%12\relax}%
\providecommand \@@startlink[1]{}%
\providecommand \@@endlink[0]{}%
\providecommand \url  [0]{\begingroup\@sanitize@url \@url }%
\providecommand \@url [1]{\endgroup\@href {#1}{\urlprefix }}%
\providecommand \urlprefix  [0]{URL }%
\providecommand \Eprint [0]{\href }%
\providecommand \doibase [0]{https://doi.org/}%
\providecommand \selectlanguage [0]{\@gobble}%
\providecommand \bibinfo  [0]{\@secondoftwo}%
\providecommand \bibfield  [0]{\@secondoftwo}%
\providecommand \translation [1]{[#1]}%
\providecommand \BibitemOpen [0]{}%
\providecommand \bibitemStop [0]{}%
\providecommand \bibitemNoStop [0]{.\EOS\space}%
\providecommand \EOS [0]{\spacefactor3000\relax}%
\providecommand \BibitemShut  [1]{\csname bibitem#1\endcsname}%
\let\auto@bib@innerbib\@empty
\bibitem [{\citenamefont {Kimble}(2008)}]{kimble2008quantum}%
  \BibitemOpen
  \bibfield  {author} {\bibinfo {author} {\bibfnamefont {H.~J.}\ \bibnamefont {Kimble}},\ }\bibfield  {title} {\bibinfo {title} {The quantum internet},\ }\href {https://doi.org/10.1038/nature07127} {\bibfield  {journal} {\bibinfo  {journal} {Nature}\ }\textbf {\bibinfo {volume} {453}},\ \bibinfo {pages} {1023} (\bibinfo {year} {2008})}\BibitemShut {NoStop}%
\bibitem [{\citenamefont {Wehner}\ \emph {et~al.}(2018)\citenamefont {Wehner}, \citenamefont {Elkouss},\ and\ \citenamefont {Hanson}}]{wehner2018quantum}%
  \BibitemOpen
  \bibfield  {author} {\bibinfo {author} {\bibfnamefont {S.}~\bibnamefont {Wehner}}, \bibinfo {author} {\bibfnamefont {D.}~\bibnamefont {Elkouss}},\ and\ \bibinfo {author} {\bibfnamefont {R.}~\bibnamefont {Hanson}},\ }\bibfield  {title} {\bibinfo {title} {Quantum internet: A vision for the road ahead},\ }\href {https://www.science.org/doi/full/10.1126/science.aam9288} {\bibfield  {journal} {\bibinfo  {journal} {Science}\ }\textbf {\bibinfo {volume} {362}},\ \bibinfo {pages} {eaam9288} (\bibinfo {year} {2018})}\BibitemShut {NoStop}%
\bibitem [{\citenamefont {Briegel}\ \emph {et~al.}(1998)\citenamefont {Briegel}, \citenamefont {D{\"u}r}, \citenamefont {Cirac},\ and\ \citenamefont {Zoller}}]{briegel1998quantum}%
  \BibitemOpen
  \bibfield  {author} {\bibinfo {author} {\bibfnamefont {H.-J.}\ \bibnamefont {Briegel}}, \bibinfo {author} {\bibfnamefont {W.}~\bibnamefont {D{\"u}r}}, \bibinfo {author} {\bibfnamefont {J.~I.}\ \bibnamefont {Cirac}},\ and\ \bibinfo {author} {\bibfnamefont {P.}~\bibnamefont {Zoller}},\ }\bibfield  {title} {\bibinfo {title} {Quantum repeaters: the role of imperfect local operations in quantum communication},\ }\href {https://doi.org/10.1103/PhysRevLett.81.5932} {\bibfield  {journal} {\bibinfo  {journal} {Physical Review Letters}\ }\textbf {\bibinfo {volume} {81}},\ \bibinfo {pages} {5932} (\bibinfo {year} {1998})}\BibitemShut {NoStop}%
\bibitem [{\citenamefont {Duan}\ \emph {et~al.}(2001)\citenamefont {Duan}, \citenamefont {Lukin}, \citenamefont {Cirac},\ and\ \citenamefont {Zoller}}]{duan2001long}%
  \BibitemOpen
  \bibfield  {author} {\bibinfo {author} {\bibfnamefont {L.-M.}\ \bibnamefont {Duan}}, \bibinfo {author} {\bibfnamefont {M.~D.}\ \bibnamefont {Lukin}}, \bibinfo {author} {\bibfnamefont {J.~I.}\ \bibnamefont {Cirac}},\ and\ \bibinfo {author} {\bibfnamefont {P.}~\bibnamefont {Zoller}},\ }\bibfield  {title} {\bibinfo {title} {Long-distance quantum communication with atomic ensembles and linear optics},\ }\href {https://www.nature.com/articles/35106500} {\bibfield  {journal} {\bibinfo  {journal} {Nature}\ }\textbf {\bibinfo {volume} {414}},\ \bibinfo {pages} {413} (\bibinfo {year} {2001})}\BibitemShut {NoStop}%
\bibitem [{\citenamefont {Simon}\ \emph {et~al.}(2007)\citenamefont {Simon}, \citenamefont {De~Riedmatten}, \citenamefont {Afzelius}, \citenamefont {Sangouard}, \citenamefont {Zbinden},\ and\ \citenamefont {Gisin}}]{simon2007quantum}%
  \BibitemOpen
  \bibfield  {author} {\bibinfo {author} {\bibfnamefont {C.}~\bibnamefont {Simon}}, \bibinfo {author} {\bibfnamefont {H.}~\bibnamefont {De~Riedmatten}}, \bibinfo {author} {\bibfnamefont {M.}~\bibnamefont {Afzelius}}, \bibinfo {author} {\bibfnamefont {N.}~\bibnamefont {Sangouard}}, \bibinfo {author} {\bibfnamefont {H.}~\bibnamefont {Zbinden}},\ and\ \bibinfo {author} {\bibfnamefont {N.}~\bibnamefont {Gisin}},\ }\bibfield  {title} {\bibinfo {title} {Quantum repeaters with photon pair sources and multimode memories},\ }\href {https://doi.org/10.1103/PhysRevLett.98.190503} {\bibfield  {journal} {\bibinfo  {journal} {Physical review letters}\ }\textbf {\bibinfo {volume} {98}},\ \bibinfo {pages} {190503} (\bibinfo {year} {2007})}\BibitemShut {NoStop}%
\bibitem [{\citenamefont {Muralidharan}\ \emph {et~al.}(2016)\citenamefont {Muralidharan}, \citenamefont {Li}, \citenamefont {Kim}, \citenamefont {L{\"u}tkenhaus}, \citenamefont {Lukin},\ and\ \citenamefont {Jiang}}]{muralidharan2016optimal}%
  \BibitemOpen
  \bibfield  {author} {\bibinfo {author} {\bibfnamefont {S.}~\bibnamefont {Muralidharan}}, \bibinfo {author} {\bibfnamefont {L.}~\bibnamefont {Li}}, \bibinfo {author} {\bibfnamefont {J.}~\bibnamefont {Kim}}, \bibinfo {author} {\bibfnamefont {N.}~\bibnamefont {L{\"u}tkenhaus}}, \bibinfo {author} {\bibfnamefont {M.~D.}\ \bibnamefont {Lukin}},\ and\ \bibinfo {author} {\bibfnamefont {L.}~\bibnamefont {Jiang}},\ }\bibfield  {title} {\bibinfo {title} {Optimal architectures for long distance quantum communication},\ }\href {https://www.nature.com/articles/srep20463} {\bibfield  {journal} {\bibinfo  {journal} {Scientific reports}\ }\textbf {\bibinfo {volume} {6}},\ \bibinfo {pages} {20463} (\bibinfo {year} {2016})}\BibitemShut {NoStop}%
\bibitem [{\citenamefont {Azuma}\ \emph {et~al.}(2023)\citenamefont {Azuma}, \citenamefont {Economou}, \citenamefont {Elkouss}, \citenamefont {Hilaire}, \citenamefont {Jiang}, \citenamefont {Lo},\ and\ \citenamefont {Tzitrin}}]{azuma2023quantum}%
  \BibitemOpen
  \bibfield  {author} {\bibinfo {author} {\bibfnamefont {K.}~\bibnamefont {Azuma}}, \bibinfo {author} {\bibfnamefont {S.~E.}\ \bibnamefont {Economou}}, \bibinfo {author} {\bibfnamefont {D.}~\bibnamefont {Elkouss}}, \bibinfo {author} {\bibfnamefont {P.}~\bibnamefont {Hilaire}}, \bibinfo {author} {\bibfnamefont {L.}~\bibnamefont {Jiang}}, \bibinfo {author} {\bibfnamefont {H.-K.}\ \bibnamefont {Lo}},\ and\ \bibinfo {author} {\bibfnamefont {I.}~\bibnamefont {Tzitrin}},\ }\bibfield  {title} {\bibinfo {title} {Quantum repeaters: From quantum networks to the quantum internet},\ }\href {https://doi.org/10.1103/RevModPhys.95.045006} {\bibfield  {journal} {\bibinfo  {journal} {Reviews of Modern Physics}\ }\textbf {\bibinfo {volume} {95}},\ \bibinfo {pages} {045006} (\bibinfo {year} {2023})}\BibitemShut {NoStop}%
\bibitem [{\citenamefont {Lei}\ \emph {et~al.}(2023)\citenamefont {Lei}, \citenamefont {Asadi}, \citenamefont {Zhong}, \citenamefont {Kuzmich}, \citenamefont {Simon},\ and\ \citenamefont {Hosseini}}]{lei2023quantum}%
  \BibitemOpen
  \bibfield  {author} {\bibinfo {author} {\bibfnamefont {Y.}~\bibnamefont {Lei}}, \bibinfo {author} {\bibfnamefont {F.~K.}\ \bibnamefont {Asadi}}, \bibinfo {author} {\bibfnamefont {T.}~\bibnamefont {Zhong}}, \bibinfo {author} {\bibfnamefont {A.}~\bibnamefont {Kuzmich}}, \bibinfo {author} {\bibfnamefont {C.}~\bibnamefont {Simon}},\ and\ \bibinfo {author} {\bibfnamefont {M.}~\bibnamefont {Hosseini}},\ }\bibfield  {title} {\bibinfo {title} {Quantum optical memory for entanglement distribution},\ }\href {https://doi.org/10.1364/OPTICA.493732} {\bibfield  {journal} {\bibinfo  {journal} {Optica}\ }\textbf {\bibinfo {volume} {10}},\ \bibinfo {pages} {1511} (\bibinfo {year} {2023})}\BibitemShut {NoStop}%
\bibitem [{\citenamefont {Zhong}\ \emph {et~al.}(2015)\citenamefont {Zhong}, \citenamefont {Hedges}, \citenamefont {Ahlefeldt}, \citenamefont {Bartholomew}, \citenamefont {Beavan}, \citenamefont {Wittig}, \citenamefont {Longdell},\ and\ \citenamefont {Sellars}}]{zhong2015optically}%
  \BibitemOpen
  \bibfield  {author} {\bibinfo {author} {\bibfnamefont {M.}~\bibnamefont {Zhong}}, \bibinfo {author} {\bibfnamefont {M.~P.}\ \bibnamefont {Hedges}}, \bibinfo {author} {\bibfnamefont {R.~L.}\ \bibnamefont {Ahlefeldt}}, \bibinfo {author} {\bibfnamefont {J.~G.}\ \bibnamefont {Bartholomew}}, \bibinfo {author} {\bibfnamefont {S.~E.}\ \bibnamefont {Beavan}}, \bibinfo {author} {\bibfnamefont {S.~M.}\ \bibnamefont {Wittig}}, \bibinfo {author} {\bibfnamefont {J.~J.}\ \bibnamefont {Longdell}},\ and\ \bibinfo {author} {\bibfnamefont {M.~J.}\ \bibnamefont {Sellars}},\ }\bibfield  {title} {\bibinfo {title} {Optically addressable nuclear spins in a solid with a six-hour coherence time},\ }\href {https://www.nature.com/articles/nature14025} {\bibfield  {journal} {\bibinfo  {journal} {Nature}\ }\textbf {\bibinfo {volume} {517}},\ \bibinfo {pages} {177} (\bibinfo {year} {2015})}\BibitemShut {NoStop}%
\bibitem [{\citenamefont {Zhou}\ \emph {et~al.}(2023)\citenamefont {Zhou}, \citenamefont {Liu}, \citenamefont {Li}, \citenamefont {Guo}, \citenamefont {Oblak}, \citenamefont {Lei}, \citenamefont {Faraon}, \citenamefont {Mazzera},\ and\ \citenamefont {de~Riedmatten}}]{zhou2023photonic}%
  \BibitemOpen
  \bibfield  {author} {\bibinfo {author} {\bibfnamefont {Z.-Q.}\ \bibnamefont {Zhou}}, \bibinfo {author} {\bibfnamefont {C.}~\bibnamefont {Liu}}, \bibinfo {author} {\bibfnamefont {C.-F.}\ \bibnamefont {Li}}, \bibinfo {author} {\bibfnamefont {G.-C.}\ \bibnamefont {Guo}}, \bibinfo {author} {\bibfnamefont {D.}~\bibnamefont {Oblak}}, \bibinfo {author} {\bibfnamefont {M.}~\bibnamefont {Lei}}, \bibinfo {author} {\bibfnamefont {A.}~\bibnamefont {Faraon}}, \bibinfo {author} {\bibfnamefont {M.}~\bibnamefont {Mazzera}},\ and\ \bibinfo {author} {\bibfnamefont {H.}~\bibnamefont {de~Riedmatten}},\ }\bibfield  {title} {\bibinfo {title} {Photonic integrated quantum memory in rare-earth doped solids},\ }\href {https://doi.org/10.1002/lpor.202300257} {\bibfield  {journal} {\bibinfo  {journal} {Laser \& Photonics Reviews}\ }\textbf {\bibinfo {volume} {17}},\ \bibinfo {pages} {2300257} (\bibinfo {year} {2023})}\BibitemShut {NoStop}%
\bibitem [{\citenamefont {Bluvstein}\ \emph {et~al.}(2024)\citenamefont {Bluvstein}, \citenamefont {Evered}, \citenamefont {Geim}, \citenamefont {Li}, \citenamefont {Zhou}, \citenamefont {Manovitz}, \citenamefont {Ebadi}, \citenamefont {Cain}, \citenamefont {Kalinowski}, \citenamefont {Hangleiter} \emph {et~al.}}]{bluvstein2024logical}%
  \BibitemOpen
  \bibfield  {author} {\bibinfo {author} {\bibfnamefont {D.}~\bibnamefont {Bluvstein}}, \bibinfo {author} {\bibfnamefont {S.~J.}\ \bibnamefont {Evered}}, \bibinfo {author} {\bibfnamefont {A.~A.}\ \bibnamefont {Geim}}, \bibinfo {author} {\bibfnamefont {S.~H.}\ \bibnamefont {Li}}, \bibinfo {author} {\bibfnamefont {H.}~\bibnamefont {Zhou}}, \bibinfo {author} {\bibfnamefont {T.}~\bibnamefont {Manovitz}}, \bibinfo {author} {\bibfnamefont {S.}~\bibnamefont {Ebadi}}, \bibinfo {author} {\bibfnamefont {M.}~\bibnamefont {Cain}}, \bibinfo {author} {\bibfnamefont {M.}~\bibnamefont {Kalinowski}}, \bibinfo {author} {\bibfnamefont {D.}~\bibnamefont {Hangleiter}}, \emph {et~al.},\ }\bibfield  {title} {\bibinfo {title} {Logical quantum processor based on reconfigurable atom arrays},\ }\href {https://doi.org/10.1038/s41586-023-06927-3} {\bibfield  {journal} {\bibinfo  {journal} {Nature}\ }\textbf {\bibinfo {volume} {626}},\ \bibinfo {pages} {58} (\bibinfo {year} {2024})}\BibitemShut {NoStop}%
\bibitem [{\citenamefont {Gu}\ \emph {et~al.}(2024)\citenamefont {Gu}, \citenamefont {Menon}, \citenamefont {Maier}, \citenamefont {Das}, \citenamefont {Chakraborty}, \citenamefont {Tittel}, \citenamefont {Bernien},\ and\ \citenamefont {Borregaard}}]{gu2024hybrid}%
  \BibitemOpen
  \bibfield  {author} {\bibinfo {author} {\bibfnamefont {F.}~\bibnamefont {Gu}}, \bibinfo {author} {\bibfnamefont {S.~G.}\ \bibnamefont {Menon}}, \bibinfo {author} {\bibfnamefont {D.}~\bibnamefont {Maier}}, \bibinfo {author} {\bibfnamefont {A.}~\bibnamefont {Das}}, \bibinfo {author} {\bibfnamefont {T.}~\bibnamefont {Chakraborty}}, \bibinfo {author} {\bibfnamefont {W.}~\bibnamefont {Tittel}}, \bibinfo {author} {\bibfnamefont {H.}~\bibnamefont {Bernien}},\ and\ \bibinfo {author} {\bibfnamefont {J.}~\bibnamefont {Borregaard}},\ }\bibfield  {title} {\bibinfo {title} {Hybrid quantum repeaters with ensemble-based quantum memories and single-spin photon transducers},\ }\href {https://doi.org/10.48550/arXiv.2401.12395} {\bibfield  {journal} {\bibinfo  {journal} {arXiv preprint arXiv:2401.12395}\ } (\bibinfo {year} {2024})}\BibitemShut {NoStop}%
\bibitem [{\citenamefont {Chaneli{\`e}re}\ \emph {et~al.}(2010)\citenamefont {Chaneli{\`e}re}, \citenamefont {Ruggiero}, \citenamefont {Bonarota}, \citenamefont {Afzelius},\ and\ \citenamefont {Le~Gou{\"e}t}}]{chaneliere2010efficient}%
  \BibitemOpen
  \bibfield  {author} {\bibinfo {author} {\bibfnamefont {T.}~\bibnamefont {Chaneli{\`e}re}}, \bibinfo {author} {\bibfnamefont {J.}~\bibnamefont {Ruggiero}}, \bibinfo {author} {\bibfnamefont {M.}~\bibnamefont {Bonarota}}, \bibinfo {author} {\bibfnamefont {M.}~\bibnamefont {Afzelius}},\ and\ \bibinfo {author} {\bibfnamefont {J.}~\bibnamefont {Le~Gou{\"e}t}},\ }\bibfield  {title} {\bibinfo {title} {Efficient light storage in a crystal using an atomic frequency comb},\ }\href {https://iopscience.iop.org/article/10.1088/1367-2630/12/2/023025} {\bibfield  {journal} {\bibinfo  {journal} {New Journal of Physics}\ }\textbf {\bibinfo {volume} {12}},\ \bibinfo {pages} {023025} (\bibinfo {year} {2010})}\BibitemShut {NoStop}%
\bibitem [{\citenamefont {Davidson}\ \emph {et~al.}(2020)\citenamefont {Davidson}, \citenamefont {Lefebvre}, \citenamefont {Zhang}, \citenamefont {Oblak},\ and\ \citenamefont {Tittel}}]{davidson2020improved}%
  \BibitemOpen
  \bibfield  {author} {\bibinfo {author} {\bibfnamefont {J.~H.}\ \bibnamefont {Davidson}}, \bibinfo {author} {\bibfnamefont {P.}~\bibnamefont {Lefebvre}}, \bibinfo {author} {\bibfnamefont {J.}~\bibnamefont {Zhang}}, \bibinfo {author} {\bibfnamefont {D.}~\bibnamefont {Oblak}},\ and\ \bibinfo {author} {\bibfnamefont {W.}~\bibnamefont {Tittel}},\ }\bibfield  {title} {\bibinfo {title} {Improved light-matter interaction for storage of quantum states of light in a thulium-doped crystal cavity},\ }\href {https://doi.org/10.1103/PhysRevA.101.042333} {\bibfield  {journal} {\bibinfo  {journal} {Physical Review A}\ }\textbf {\bibinfo {volume} {101}},\ \bibinfo {pages} {042333} (\bibinfo {year} {2020})}\BibitemShut {NoStop}%
\bibitem [{\citenamefont {Sinclair}\ \emph {et~al.}(2014)\citenamefont {Sinclair}, \citenamefont {Saglamyurek}, \citenamefont {Mallahzadeh}, \citenamefont {Slater}, \citenamefont {George}, \citenamefont {Ricken}, \citenamefont {Hedges}, \citenamefont {Oblak}, \citenamefont {Simon}, \citenamefont {Sohler} \emph {et~al.}}]{sinclair2014spectral}%
  \BibitemOpen
  \bibfield  {author} {\bibinfo {author} {\bibfnamefont {N.}~\bibnamefont {Sinclair}}, \bibinfo {author} {\bibfnamefont {E.}~\bibnamefont {Saglamyurek}}, \bibinfo {author} {\bibfnamefont {H.}~\bibnamefont {Mallahzadeh}}, \bibinfo {author} {\bibfnamefont {J.~A.}\ \bibnamefont {Slater}}, \bibinfo {author} {\bibfnamefont {M.}~\bibnamefont {George}}, \bibinfo {author} {\bibfnamefont {R.}~\bibnamefont {Ricken}}, \bibinfo {author} {\bibfnamefont {M.~P.}\ \bibnamefont {Hedges}}, \bibinfo {author} {\bibfnamefont {D.}~\bibnamefont {Oblak}}, \bibinfo {author} {\bibfnamefont {C.}~\bibnamefont {Simon}}, \bibinfo {author} {\bibfnamefont {W.}~\bibnamefont {Sohler}}, \emph {et~al.},\ }\bibfield  {title} {\bibinfo {title} {Spectral multiplexing for scalable quantum photonics using an atomic frequency comb quantum memory and feed-forward control},\ }\href {https://doi.org/10.1103/PhysRevLett.113.053603} {\bibfield  {journal} {\bibinfo  {journal} {Physical review letters}\ }\textbf {\bibinfo {volume} {113}},\ \bibinfo {pages}
  {053603} (\bibinfo {year} {2014})}\BibitemShut {NoStop}%
\bibitem [{\citenamefont {Askarani}\ \emph {et~al.}(2020)\citenamefont {Askarani}, \citenamefont {Davidson}, \citenamefont {Verma}, \citenamefont {Shaw}, \citenamefont {Nam}, \citenamefont {Lutz}, \citenamefont {Amaral}, \citenamefont {Oblak}, \citenamefont {Tittel} \emph {et~al.}}]{askarani2020entanglement}%
  \BibitemOpen
  \bibfield  {author} {\bibinfo {author} {\bibfnamefont {M.~F.}\ \bibnamefont {Askarani}}, \bibinfo {author} {\bibfnamefont {J.~H.}\ \bibnamefont {Davidson}}, \bibinfo {author} {\bibfnamefont {V.~B.}\ \bibnamefont {Verma}}, \bibinfo {author} {\bibfnamefont {M.~D.}\ \bibnamefont {Shaw}}, \bibinfo {author} {\bibfnamefont {S.~W.}\ \bibnamefont {Nam}}, \bibinfo {author} {\bibfnamefont {T.}~\bibnamefont {Lutz}}, \bibinfo {author} {\bibfnamefont {G.~C.}\ \bibnamefont {Amaral}}, \bibinfo {author} {\bibfnamefont {D.}~\bibnamefont {Oblak}}, \bibinfo {author} {\bibfnamefont {W.}~\bibnamefont {Tittel}}, \emph {et~al.},\ }\bibfield  {title} {\bibinfo {title} {Entanglement and nonlocality between disparate solid-state quantum memories mediated by photons},\ }\href {https://doi.org/10.1103/PhysRevResearch.2.013039} {\bibfield  {journal} {\bibinfo  {journal} {Physical Review Research}\ }\textbf {\bibinfo {volume} {2}},\ \bibinfo {pages} {013039} (\bibinfo {year} {2020})}\BibitemShut {NoStop}%
\bibitem [{\citenamefont {Thiel}\ \emph {et~al.}(2014)\citenamefont {Thiel}, \citenamefont {Sinclair}, \citenamefont {Tittel},\ and\ \citenamefont {Cone}}]{thiel2014tm}%
  \BibitemOpen
  \bibfield  {author} {\bibinfo {author} {\bibfnamefont {C.~W.}\ \bibnamefont {Thiel}}, \bibinfo {author} {\bibfnamefont {N.}~\bibnamefont {Sinclair}}, \bibinfo {author} {\bibfnamefont {W.}~\bibnamefont {Tittel}},\ and\ \bibinfo {author} {\bibfnamefont {R.~L.}\ \bibnamefont {Cone}},\ }\bibfield  {title} {\bibinfo {title} {Tm materials for spectrally multiplexed quantum memories},\ }\href {https://doi.org/10.1103/PhysRevLett.113.160501} {\bibfield  {journal} {\bibinfo  {journal} {Physical review letters}\ }\textbf {\bibinfo {volume} {113}},\ \bibinfo {pages} {160501} (\bibinfo {year} {2014})}\BibitemShut {NoStop}%
\bibitem [{\citenamefont {Askarani}\ \emph {et~al.}(2021)\citenamefont {Askarani}, \citenamefont {Das}, \citenamefont {Davidson}, \citenamefont {Amaral}, \citenamefont {Sinclair}, \citenamefont {Slater}, \citenamefont {Marzban}, \citenamefont {Thiel}, \citenamefont {Cone}, \citenamefont {Oblak} \emph {et~al.}}]{askarani2021long}%
  \BibitemOpen
  \bibfield  {author} {\bibinfo {author} {\bibfnamefont {M.~F.}\ \bibnamefont {Askarani}}, \bibinfo {author} {\bibfnamefont {A.}~\bibnamefont {Das}}, \bibinfo {author} {\bibfnamefont {J.~H.}\ \bibnamefont {Davidson}}, \bibinfo {author} {\bibfnamefont {G.~C.}\ \bibnamefont {Amaral}}, \bibinfo {author} {\bibfnamefont {N.}~\bibnamefont {Sinclair}}, \bibinfo {author} {\bibfnamefont {J.~A.}\ \bibnamefont {Slater}}, \bibinfo {author} {\bibfnamefont {S.}~\bibnamefont {Marzban}}, \bibinfo {author} {\bibfnamefont {C.~W.}\ \bibnamefont {Thiel}}, \bibinfo {author} {\bibfnamefont {R.~L.}\ \bibnamefont {Cone}}, \bibinfo {author} {\bibfnamefont {D.}~\bibnamefont {Oblak}}, \emph {et~al.},\ }\bibfield  {title} {\bibinfo {title} {Long-lived solid-state optical memory for high-rate quantum repeaters},\ }\href {https://doi.org/10.1103/PhysRevLett.127.220502} {\bibfield  {journal} {\bibinfo  {journal} {Physical review letters}\ }\textbf {\bibinfo {volume} {127}},\ \bibinfo {pages} {220502} (\bibinfo {year} {2021})}\BibitemShut
  {NoStop}%
\bibitem [{\citenamefont {Dutta}\ \emph {et~al.}(2023)\citenamefont {Dutta}, \citenamefont {Zhao}, \citenamefont {Saha}, \citenamefont {Farfurnik}, \citenamefont {Goldschmidt},\ and\ \citenamefont {Waks}}]{dutta2023atomic}%
  \BibitemOpen
  \bibfield  {author} {\bibinfo {author} {\bibfnamefont {S.}~\bibnamefont {Dutta}}, \bibinfo {author} {\bibfnamefont {Y.}~\bibnamefont {Zhao}}, \bibinfo {author} {\bibfnamefont {U.}~\bibnamefont {Saha}}, \bibinfo {author} {\bibfnamefont {D.}~\bibnamefont {Farfurnik}}, \bibinfo {author} {\bibfnamefont {E.~A.}\ \bibnamefont {Goldschmidt}},\ and\ \bibinfo {author} {\bibfnamefont {E.}~\bibnamefont {Waks}},\ }\bibfield  {title} {\bibinfo {title} {An atomic frequency comb memory in rare-earth-doped thin-film lithium niobate},\ }\href {https://doi.org/10.1021/acsphotonics.2c01835} {\bibfield  {journal} {\bibinfo  {journal} {ACS Photonics}\ }\textbf {\bibinfo {volume} {10}},\ \bibinfo {pages} {1104} (\bibinfo {year} {2023})}\BibitemShut {NoStop}%
\bibitem [{\citenamefont {Zhao}\ \emph {et~al.}(2024)\citenamefont {Zhao}, \citenamefont {Renaud}, \citenamefont {Farfurnik}, \citenamefont {Jiang}, \citenamefont {Dutta}, \citenamefont {Sinclair}, \citenamefont {Lon{\v{c}}ar},\ and\ \citenamefont {Waks}}]{zhao2024cavity}%
  \BibitemOpen
  \bibfield  {author} {\bibinfo {author} {\bibfnamefont {Y.}~\bibnamefont {Zhao}}, \bibinfo {author} {\bibfnamefont {D.}~\bibnamefont {Renaud}}, \bibinfo {author} {\bibfnamefont {D.}~\bibnamefont {Farfurnik}}, \bibinfo {author} {\bibfnamefont {Y.}~\bibnamefont {Jiang}}, \bibinfo {author} {\bibfnamefont {S.}~\bibnamefont {Dutta}}, \bibinfo {author} {\bibfnamefont {N.}~\bibnamefont {Sinclair}}, \bibinfo {author} {\bibfnamefont {M.}~\bibnamefont {Lon{\v{c}}ar}},\ and\ \bibinfo {author} {\bibfnamefont {E.}~\bibnamefont {Waks}},\ }\bibfield  {title} {\bibinfo {title} {Cavity-enhanced narrowband spectral filters using rare-earth ions doped in thin-film lithium niobate},\ }\href {https://doi.org/10.1038/s44310-024-00023-8} {\bibfield  {journal} {\bibinfo  {journal} {npj Nanophotonics}\ }\textbf {\bibinfo {volume} {1}},\ \bibinfo {pages} {22} (\bibinfo {year} {2024})}\BibitemShut {NoStop}%
\bibitem [{\citenamefont {Pak}\ \emph {et~al.}(2022)\citenamefont {Pak}, \citenamefont {Nandi}, \citenamefont {Titze}, \citenamefont {Bielejec}, \citenamefont {Alaeian},\ and\ \citenamefont {Hosseini}}]{pak2022long}%
  \BibitemOpen
  \bibfield  {author} {\bibinfo {author} {\bibfnamefont {D.}~\bibnamefont {Pak}}, \bibinfo {author} {\bibfnamefont {A.}~\bibnamefont {Nandi}}, \bibinfo {author} {\bibfnamefont {M.}~\bibnamefont {Titze}}, \bibinfo {author} {\bibfnamefont {E.~S.}\ \bibnamefont {Bielejec}}, \bibinfo {author} {\bibfnamefont {H.}~\bibnamefont {Alaeian}},\ and\ \bibinfo {author} {\bibfnamefont {M.}~\bibnamefont {Hosseini}},\ }\bibfield  {title} {\bibinfo {title} {Long-range cooperative resonances in rare-earth ion arrays inside photonic resonators},\ }\href {https://doi.org/10.1038/s42005-022-00871-w} {\bibfield  {journal} {\bibinfo  {journal} {Communications Physics}\ }\textbf {\bibinfo {volume} {5}},\ \bibinfo {pages} {89} (\bibinfo {year} {2022})}\BibitemShut {NoStop}%
\bibitem [{\citenamefont {Louchet}\ \emph {et~al.}(2007)\citenamefont {Louchet}, \citenamefont {Habib}, \citenamefont {Crozatier}, \citenamefont {Lorger{\'e}}, \citenamefont {Goldfarb}, \citenamefont {Bretenaker}, \citenamefont {Le~Gou{\"e}t}, \citenamefont {Guillot-No{\"e}l},\ and\ \citenamefont {Goldner}}]{louchet2007branching}%
  \BibitemOpen
  \bibfield  {author} {\bibinfo {author} {\bibfnamefont {A.}~\bibnamefont {Louchet}}, \bibinfo {author} {\bibfnamefont {J.~S.}\ \bibnamefont {Habib}}, \bibinfo {author} {\bibfnamefont {V.}~\bibnamefont {Crozatier}}, \bibinfo {author} {\bibfnamefont {I.}~\bibnamefont {Lorger{\'e}}}, \bibinfo {author} {\bibfnamefont {F.}~\bibnamefont {Goldfarb}}, \bibinfo {author} {\bibfnamefont {F.}~\bibnamefont {Bretenaker}}, \bibinfo {author} {\bibfnamefont {J.-L.}\ \bibnamefont {Le~Gou{\"e}t}}, \bibinfo {author} {\bibfnamefont {O.}~\bibnamefont {Guillot-No{\"e}l}},\ and\ \bibinfo {author} {\bibfnamefont {P.}~\bibnamefont {Goldner}},\ }\bibfield  {title} {\bibinfo {title} {Branching ratio measurement of a $\lambda$ system in tm 3+: Yag under a magnetic field},\ }\href {https://doi.org/10.1103/PhysRevB.75.035131} {\bibfield  {journal} {\bibinfo  {journal} {Physical Review B}\ }\textbf {\bibinfo {volume} {75}},\ \bibinfo {pages} {035131} (\bibinfo {year} {2007})}\BibitemShut {NoStop}%
\bibitem [{\citenamefont {Lei}\ \emph {et~al.}(2024)\citenamefont {Lei}, \citenamefont {An}, \citenamefont {Li},\ and\ \citenamefont {Hosseini}}]{lei2024algorithmic}%
  \BibitemOpen
  \bibfield  {author} {\bibinfo {author} {\bibfnamefont {Y.}~\bibnamefont {Lei}}, \bibinfo {author} {\bibfnamefont {H.}~\bibnamefont {An}}, \bibinfo {author} {\bibfnamefont {Z.}~\bibnamefont {Li}},\ and\ \bibinfo {author} {\bibfnamefont {M.}~\bibnamefont {Hosseini}},\ }\bibfield  {title} {\bibinfo {title} {Algorithmic optimization of quantum optical storage in solids},\ }\href {https://doi.org/10.1103/PhysRevResearch.6.033153} {\bibfield  {journal} {\bibinfo  {journal} {Physical Review Research}\ }\textbf {\bibinfo {volume} {6}},\ \bibinfo {pages} {033153} (\bibinfo {year} {2024})}\BibitemShut {NoStop}%
\bibitem [{\citenamefont {Black}(2001)}]{black2001introduction}%
  \BibitemOpen
  \bibfield  {author} {\bibinfo {author} {\bibfnamefont {E.~D.}\ \bibnamefont {Black}},\ }\bibfield  {title} {\bibinfo {title} {An introduction to pound--drever--hall laser frequency stabilization},\ }\href {https://doi.org/10.1119/1.1286663} {\bibfield  {journal} {\bibinfo  {journal} {American journal of physics}\ }\textbf {\bibinfo {volume} {69}},\ \bibinfo {pages} {79} (\bibinfo {year} {2001})}\BibitemShut {NoStop}%
\bibitem [{\citenamefont {Sinclair}\ \emph {et~al.}(2021)\citenamefont {Sinclair}, \citenamefont {Oblak}, \citenamefont {Saglamyurek}, \citenamefont {Cone}, \citenamefont {Thiel},\ and\ \citenamefont {Tittel}}]{sinclair2021optical}%
  \BibitemOpen
  \bibfield  {author} {\bibinfo {author} {\bibfnamefont {N.}~\bibnamefont {Sinclair}}, \bibinfo {author} {\bibfnamefont {D.}~\bibnamefont {Oblak}}, \bibinfo {author} {\bibfnamefont {E.}~\bibnamefont {Saglamyurek}}, \bibinfo {author} {\bibfnamefont {R.~L.}\ \bibnamefont {Cone}}, \bibinfo {author} {\bibfnamefont {C.~W.}\ \bibnamefont {Thiel}},\ and\ \bibinfo {author} {\bibfnamefont {W.}~\bibnamefont {Tittel}},\ }\bibfield  {title} {\bibinfo {title} {Optical coherence and energy-level properties of a tm 3+-doped li nb o 3 waveguide at subkelvin temperatures},\ }\href {https://doi.org/10.1103/PhysRevB.103.134105} {\bibfield  {journal} {\bibinfo  {journal} {Physical Review B}\ }\textbf {\bibinfo {volume} {103}},\ \bibinfo {pages} {134105} (\bibinfo {year} {2021})}\BibitemShut {NoStop}%
\bibitem [{\citenamefont {De~Seze}\ \emph {et~al.}(2006)\citenamefont {De~Seze}, \citenamefont {Louchet}, \citenamefont {Crozatier}, \citenamefont {Lorger{\'e}}, \citenamefont {Bretenaker}, \citenamefont {Le~Gou{\"e}t}, \citenamefont {Guillot-No{\"e}l},\ and\ \citenamefont {Goldner}}]{de2006experimental}%
  \BibitemOpen
  \bibfield  {author} {\bibinfo {author} {\bibfnamefont {F.}~\bibnamefont {De~Seze}}, \bibinfo {author} {\bibfnamefont {A.}~\bibnamefont {Louchet}}, \bibinfo {author} {\bibfnamefont {V.}~\bibnamefont {Crozatier}}, \bibinfo {author} {\bibfnamefont {I.}~\bibnamefont {Lorger{\'e}}}, \bibinfo {author} {\bibfnamefont {F.}~\bibnamefont {Bretenaker}}, \bibinfo {author} {\bibfnamefont {J.-L.}\ \bibnamefont {Le~Gou{\"e}t}}, \bibinfo {author} {\bibfnamefont {O.}~\bibnamefont {Guillot-No{\"e}l}},\ and\ \bibinfo {author} {\bibfnamefont {P.}~\bibnamefont {Goldner}},\ }\bibfield  {title} {\bibinfo {title} {Experimental tailoring of a three-level $\lambda$ system in tm 3+: Yag},\ }\href {https://doi.org/10.1103/PhysRevB.73.085112} {\bibfield  {journal} {\bibinfo  {journal} {Physical Review B—Condensed Matter and Materials Physics}\ }\textbf {\bibinfo {volume} {73}},\ \bibinfo {pages} {085112} (\bibinfo {year} {2006})}\BibitemShut {NoStop}%
\bibitem [{\citenamefont {Afzelius}\ \emph {et~al.}(2009)\citenamefont {Afzelius}, \citenamefont {Simon}, \citenamefont {De~Riedmatten},\ and\ \citenamefont {Gisin}}]{afzelius2009multimode}%
  \BibitemOpen
  \bibfield  {author} {\bibinfo {author} {\bibfnamefont {M.}~\bibnamefont {Afzelius}}, \bibinfo {author} {\bibfnamefont {C.}~\bibnamefont {Simon}}, \bibinfo {author} {\bibfnamefont {H.}~\bibnamefont {De~Riedmatten}},\ and\ \bibinfo {author} {\bibfnamefont {N.}~\bibnamefont {Gisin}},\ }\bibfield  {title} {\bibinfo {title} {Multimode quantum memory based on atomic frequency combs},\ }\href {https://journals.aps.org/pra/abstract/10.1103/PhysRevA.79.052329} {\bibfield  {journal} {\bibinfo  {journal} {Physical Review A}\ }\textbf {\bibinfo {volume} {79}},\ \bibinfo {pages} {052329} (\bibinfo {year} {2009})}\BibitemShut {NoStop}%
\bibitem [{\citenamefont {Bonarota}\ \emph {et~al.}(2010)\citenamefont {Bonarota}, \citenamefont {Ruggiero}, \citenamefont {Le~Gou{\"e}t},\ and\ \citenamefont {Chaneli{\`e}re}}]{bonarota2010efficiency}%
  \BibitemOpen
  \bibfield  {author} {\bibinfo {author} {\bibfnamefont {M.}~\bibnamefont {Bonarota}}, \bibinfo {author} {\bibfnamefont {J.}~\bibnamefont {Ruggiero}}, \bibinfo {author} {\bibfnamefont {J.-L.}\ \bibnamefont {Le~Gou{\"e}t}},\ and\ \bibinfo {author} {\bibfnamefont {T.}~\bibnamefont {Chaneli{\`e}re}},\ }\bibfield  {title} {\bibinfo {title} {Efficiency optimization for atomic frequency comb storage},\ }\href {https://doi.org/10.1103/PhysRevA.81.033803} {\bibfield  {journal} {\bibinfo  {journal} {Physical Review A}\ }\textbf {\bibinfo {volume} {81}},\ \bibinfo {pages} {033803} (\bibinfo {year} {2010})}\BibitemShut {NoStop}%
\bibitem [{\citenamefont {Jobez}\ \emph {et~al.}(2016)\citenamefont {Jobez}, \citenamefont {Timoney}, \citenamefont {Laplane}, \citenamefont {Etesse}, \citenamefont {Ferrier}, \citenamefont {Goldner}, \citenamefont {Gisin},\ and\ \citenamefont {Afzelius}}]{jobez2016towards}%
  \BibitemOpen
  \bibfield  {author} {\bibinfo {author} {\bibfnamefont {P.}~\bibnamefont {Jobez}}, \bibinfo {author} {\bibfnamefont {N.}~\bibnamefont {Timoney}}, \bibinfo {author} {\bibfnamefont {C.}~\bibnamefont {Laplane}}, \bibinfo {author} {\bibfnamefont {J.}~\bibnamefont {Etesse}}, \bibinfo {author} {\bibfnamefont {A.}~\bibnamefont {Ferrier}}, \bibinfo {author} {\bibfnamefont {P.}~\bibnamefont {Goldner}}, \bibinfo {author} {\bibfnamefont {N.}~\bibnamefont {Gisin}},\ and\ \bibinfo {author} {\bibfnamefont {M.}~\bibnamefont {Afzelius}},\ }\bibfield  {title} {\bibinfo {title} {Towards highly multimode optical quantum memory for quantum repeaters},\ }\href {https://doi.org/10.1103/PhysRevA.93.032327} {\bibfield  {journal} {\bibinfo  {journal} {Physical Review A}\ }\textbf {\bibinfo {volume} {93}},\ \bibinfo {pages} {032327} (\bibinfo {year} {2016})}\BibitemShut {NoStop}%
\bibitem [{Zongfeng Li, \textit{et al.}, paper in preparation()}]{Li2025}%
  \BibitemOpen
  Zongfeng Li, \textit{et al.}, paper in preparation,\ \href@noop {} {}\BibitemShut {NoStop}%
\bibitem [{\citenamefont {Yang}\ \emph {et~al.}(2018)\citenamefont {Yang}, \citenamefont {Zhou}, \citenamefont {Hua}, \citenamefont {Liu}, \citenamefont {Li}, \citenamefont {Li}, \citenamefont {Ma}, \citenamefont {Liu}, \citenamefont {Liang}, \citenamefont {Li} \emph {et~al.}}]{yang2018multiplexed}%
  \BibitemOpen
  \bibfield  {author} {\bibinfo {author} {\bibfnamefont {T.-S.}\ \bibnamefont {Yang}}, \bibinfo {author} {\bibfnamefont {Z.-Q.}\ \bibnamefont {Zhou}}, \bibinfo {author} {\bibfnamefont {Y.-L.}\ \bibnamefont {Hua}}, \bibinfo {author} {\bibfnamefont {X.}~\bibnamefont {Liu}}, \bibinfo {author} {\bibfnamefont {Z.-F.}\ \bibnamefont {Li}}, \bibinfo {author} {\bibfnamefont {P.-Y.}\ \bibnamefont {Li}}, \bibinfo {author} {\bibfnamefont {Y.}~\bibnamefont {Ma}}, \bibinfo {author} {\bibfnamefont {C.}~\bibnamefont {Liu}}, \bibinfo {author} {\bibfnamefont {P.-J.}\ \bibnamefont {Liang}}, \bibinfo {author} {\bibfnamefont {X.}~\bibnamefont {Li}}, \emph {et~al.},\ }\bibfield  {title} {\bibinfo {title} {Multiplexed storage and real-time manipulation based on a multiple degree-of-freedom quantum memory},\ }\href {https://doi.org/10.1038/s41467-018-05669-5} {\bibfield  {journal} {\bibinfo  {journal} {Nature communications}\ }\textbf {\bibinfo {volume} {9}},\ \bibinfo {pages} {3407} (\bibinfo {year} {2018})}\BibitemShut {NoStop}%
\bibitem [{\citenamefont {Li}\ \emph {et~al.}(2024)\citenamefont {Li}, \citenamefont {Lei}, \citenamefont {Kling},\ and\ \citenamefont {Hosseini}}]{li2024efficient}%
  \BibitemOpen
  \bibfield  {author} {\bibinfo {author} {\bibfnamefont {Z.}~\bibnamefont {Li}}, \bibinfo {author} {\bibfnamefont {Y.}~\bibnamefont {Lei}}, \bibinfo {author} {\bibfnamefont {T.}~\bibnamefont {Kling}},\ and\ \bibinfo {author} {\bibfnamefont {M.}~\bibnamefont {Hosseini}},\ }\bibfield  {title} {\bibinfo {title} {Efficient storage of multidimensional telecom photons in a solid-state quantum memory},\ }\href {https://doi.org/10.48550/arXiv.2412.05480} {\bibfield  {journal} {\bibinfo  {journal} {arXiv preprint arXiv:2412.05480}\ } (\bibinfo {year} {2024})}\BibitemShut {NoStop}%
\bibitem [{\citenamefont {Afzelius}\ and\ \citenamefont {Simon}(2010)}]{afzelius2010impedance}%
  \BibitemOpen
  \bibfield  {author} {\bibinfo {author} {\bibfnamefont {M.}~\bibnamefont {Afzelius}}\ and\ \bibinfo {author} {\bibfnamefont {C.}~\bibnamefont {Simon}},\ }\bibfield  {title} {\bibinfo {title} {Impedance-matched cavity quantum memory},\ }\href {https://doi.org/10.1103/PhysRevA.82.022310} {\bibfield  {journal} {\bibinfo  {journal} {Physical Review A}\ }\textbf {\bibinfo {volume} {82}},\ \bibinfo {pages} {022310} (\bibinfo {year} {2010})}\BibitemShut {NoStop}%
\bibitem [{\citenamefont {Moiseev}\ \emph {et~al.}(2010)\citenamefont {Moiseev}, \citenamefont {Andrianov},\ and\ \citenamefont {Gubaidullin}}]{moiseev2010efficient}%
  \BibitemOpen
  \bibfield  {author} {\bibinfo {author} {\bibfnamefont {S.~A.}\ \bibnamefont {Moiseev}}, \bibinfo {author} {\bibfnamefont {S.~N.}\ \bibnamefont {Andrianov}},\ and\ \bibinfo {author} {\bibfnamefont {F.~F.}\ \bibnamefont {Gubaidullin}},\ }\bibfield  {title} {\bibinfo {title} {Efficient multimode quantum memory based on photon echo in an optimal qed cavity},\ }\href {https://doi.org/10.1103/PhysRevA.82.022311} {\bibfield  {journal} {\bibinfo  {journal} {Physical review A}\ }\textbf {\bibinfo {volume} {82}},\ \bibinfo {pages} {022311} (\bibinfo {year} {2010})}\BibitemShut {NoStop}%
\end{thebibliography}%

\end{document}